\documentclass{aa}
\usepackage{graphics}

\def\beq{\begin{equation}}
\def\eeq{\end{equation}}
\def\bey{\begin{eqnarray}}
\def\eey{\end{eqnarray}}
\def\kms{\rm \,km\,s^{-1}}

\def\mpc{\,{\rm {Mpc}}}

\def\Lmin{L_{\rm min}}
\def\Lmax{L_{\rm max}}
\def\zmax{z_{\rm max}}
\def\rH{{\cal R}_\star}
\def\Lstar{L_\star}

\def\apj{ApJ}

\def\prl{Phys. Rev. Lett.}

\def\chisq{\chi^2}
\def\dL{\Delta L}
\def\ergs{\rm erg~s^{-1}}
\def\cR{{\rm ~counts~cm^{-2}~s^{-1}}}
\def\ru{{\rm Gpc^{-3}~ yr^{-1}}}

\begin{document}

\thesaurus{01         
	(13.07.1;
	11.05.2  
	11.06.1; 
	11.19.3  
	)}
\title{The Nature of the Host Galaxies for Gamma-Ray Bursts}

\author{Shude Mao \and H.J. Mo}

\offprints{S. Mao}
\mail{(smao, hom)@mpa-garching.mpg.de}

\institute{Max-Planck-Institut f\"ur Astrophysik, 
	Karl-Schwarzschild-Strasse 1, 85740 Garching, Germany}

\date{Received 1998; accepted 1998}
\titlerunning{The Nature of the Host Galaxies for Gamma-Ray Bursts}
\authorrunning{Mao \& Mo}

\maketitle

\begin{abstract}

It has been suggested recently that the rate of gamma-ray bursts
(GRBs) is proportional to the star formation rate in the 
universe. In this paper, we study the nature of GRB hosts expected
in this scenario. We improve upon previous studies by incorporating
a luminosity function for the GRBs, as required by observations. 
This model provides a good match to the observed number counts of 
GRBs as a function of peak-count rate. The model predicts that
the host galaxies have their redshift distribution peaked around
$z\sim 1$, and about 15 percent have $z>2.5$. This high-redshift 
fraction have the same properties as the star-forming galaxies
recently discovered by the Lyman-break technique. At $z \la 1$, 
many of the GRBs may be hosted by faint blue galaxies. Using a photometric 
redshift sample of galaxies from the Hubble Deep Field, we find that 
the host galaxies have magnitudes in the range from 21.5 to 28 in the 
I-band, and about 90 percent of them have semi-major axis smaller than
$1.3\arcsec$. Assuming isotropic emission, the 
typical peak-luminosity and total energy of GRBs are
$\sim 10^{51} \ergs$ and $10^{52} {\rm erg}$
in an Einstein-de Sitter universe with $H_0=100\kms\mpc^{-1}$.
We also discuss further observational tests of this  
scenario.

\end{abstract}

\section{Introduction}

The recent redshift measurements of the optical counterparts 
associated with gamma-ray bursts (GRBs) have 
established their cosmological origin (e.g. Paczy\'nski 1986). 
Although the light curves of GRB afterglows can be well
accommodated in the relativistic fireball model (e.g., Waxman 1997) 
the nature of their hosts remains a mystery. 

Before the discovery of the optical counterparts, 
the hosts were thought to be at modest redshift ($z < 1$).
This conclusion was based on the simple assumption that 
the comoving rate of GRBs is a constant (e.g., Mao \& Paczy\'{n}ski 1992; 
Piran 1992; Dermer 1992). However, there is now considerable evidence 
that the hosts of GRBs may be at substantially higher redshift, 
both from the redshift measurement for GRB 971214 ($z=3.42$, 
Kulkarni et al. 1998) and (indirectly) from the time dilations 
of GRBs (Fenimore \& Bloom 1995; Bonnell et al. 1997).
Recent attention has therefore been focused on models 
in which burst activity is linked to the massive star 
formation, as would be expected in the ``failed-supernova'' model
(Woosley 1993) or in the ``hypernova'' 
model (Paczy\'nski 1998). Totani (1997, 1998) and Wijers et al. 
(1998) have studied whether such a scenario is consistent with
the observed number counts of GRBs as a function of peak-count rate.
All these studies assumed that GRBs are standard candles 
(i.e. they have the same intrinsic luminosity) and were based on the 
star formation history given by Madau et al. (1998). 

 In this paper we examine further the star-formation origin
of gamma-ray bursts. Our approach is different from that of
earlier studies in several aspects. First, we incorporate 
a luminosity function for GRBs. The standard-candle assumption used 
in previous studies is no longer tenable because the inferred 
peak luminosities for GRB 971214 and 970508 differ
by a factor of $\sim 30$. Second, we use the star formation rate
given by Steidel et al. (1998). This rate at $z \ga 1.5$
is substantially higher than that used in the previous 
analyses, and therefore may change the predicted redshift distribution 
of GRBs. Third, we make specific predictions for the sizes 
and magnitudes of the host galaxies using the Hubble Deep Field 
data, which allow us to examine whether the model is 
consistent with the observations that GRB hosts 
are usually faint and small.
 

\section {Model}

Since most of our results are nearly independent of cosmology, 
we adopt an Einstein-de Sitter cosmology in our discussion.
The luminosity function of gamma-ray bursts is assumed to be
independent of redshift and has a power-law form
\beq
\phi(L) dL = \rH (L/\Lstar)^{\beta} d(L/\Lstar),~~\Lmin \le L \le \Lmax,
\eeq
where $\Lstar$ is a characteristic luminosity (to be chosen below).
(We also tried a log-normal distribution and found very similar 
results.)~
The rest-frame GRB spectra are also modelled as a power-law
($dN/dE \propto E^{-\alpha}$). This is clearly a
simplification given that GRBs have diverse spectra
(Band et al. 1993; Mallozzi, Pendleton \&
Paciesas 1996). A more realistic treatment involves the correction of the 
observed spectra (for bright GRBs) to the rest frame, 
because even these bright bursts may cover a substantial range
in redshift (Fenimore \& Bloom 1995). This has been performed 
for the standard-candle case by Fenimore \& Bloom (1995). 
Unfortunately, such a treatment is more complicated for our
case with a luminosity function. We therefore adopt the power-law
simplification. Mallozzi et al. (1996) gave $\alpha=1.1\pm 0.3$; we
take a slightly larger value ($\alpha=1.5$) to partially take into
account the high-energy steepening in the GRB spectra. 
As we will see in Section 3, this choice reproduces the
results obtained by other authors using more realistic spectra.

The GRB rate is taken to be proportional to the star formation
rate. We use the most recent star formation history determined
by Steidel et al. (1998). The star formation rate at $1.2<z<4$ 
given by Steidel et al. is much higher than the estimate of 
Madau et al. (1998). At the moment there is no observational 
constraint on the star formation rate for $z>4$. Beyond 
this redshift, we simply assume that the star formation rate 
drops by a factor of 10 per unit redshift.  
Since most bursts are below redshift of 4 in our model
(see Fig. 1), our results are insensitive to this extrapolation. 

The four model parameters, $\dL \equiv \log(\Lmin/\Lmax)$, $\beta$,
$\Lmax$, and the rate parameter $\rH$ are found using the same procedures 
as in Fenimore \& Bloom (1995). This method minimizes the 
$\chi^2$ measure of the observed counts of GRBs, $N$, in 11 bins
of the peak-count rate, $P$ (in units of $\cR$).
$\Lmax$ can also be substituted by the maximum redshift ($\zmax$) 
out to which a burst with luminosity $\Lmax$ can still be seen.
We choose $\Lstar$ to be $\Lmax/30$, approximately the median
peak luminosity observed. Following Fenimore \& Bloom,
we only consider bursts with $P>1$ on the 1024ms time-scale
from the BATSE instrument to avoid threshold effects. Our analysis
applies primarily to the long bursts in the BATSE catalogue and we quote
the energy in the 50-300 keV range.

\section{Results}

We first fit the $\log N-\log P$ relation assuming 
a standard-candle model for GRB luminosities
and a {\it constant} burst rate (in comoving units) in an Einstein
de-Sitter universe. This model has two parameters: $\Lmax$ ($\zmax$) and
$\rH$. The best fit model has $\zmax=0.73, \rH=45
h^3\ru$, where $h$ is the present-day Hubble constant in units of 
$100\kms\mpc^{-1}$. These values are in good agreement with those
obtained from a more sophisticated modelling by
Fenimore \& Bloom (1995) and Wijers et al. (1998). The fit is
excellent, with $\chisq=9.1$ for 9 degrees of freedom,
confirming the conclusions of previous studies
that a non-evolving standard-candle GRB population
provides a good fit to the $\log N - \log P$ curve. However, 
the standard-candle assumption is no longer supported by recent
observations, as discussed in Section 1. 

\begin{figure}
\vspace{-1cm}
\resizebox{\hsize}{!}{\includegraphics{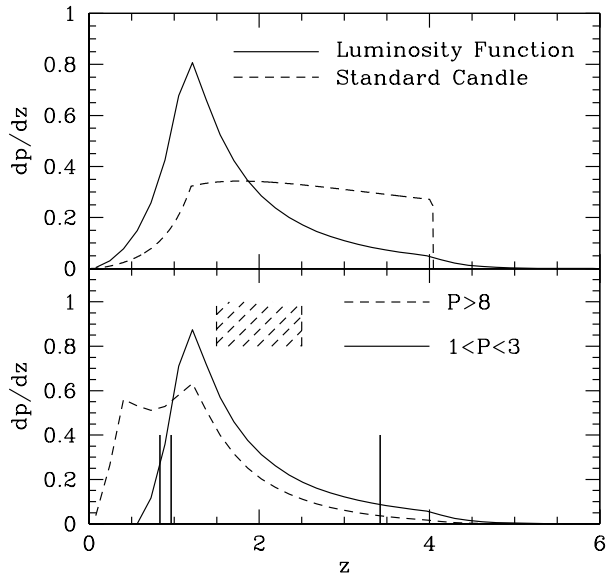}}
\vspace{-0.8cm}
\caption{
The top panel shows the predicted redshift distributions for
two models with the GRBs luminosity function modelled as a power-law 
(solid line) and standard candles (dashed line). Both models assume the
burst rate to be proportional to the star formation rate in the universe.
The bottom panel shows the redshift distribution for GRBs with
$P>8$ (dashed) and $1<P<3$ (solid), where $P$ is 
the peak-count rate in units of $\cR$. The three solid lines show the
three redshifts (from left to right) for GRB 970508, 971214 
and 980703 (see text). The shaded region is the likely redshift range
for GRB 970228. 
}
\end{figure}

 Next we fit the data with the model described in the last
section. The best-fit parameters are 
$\beta=-2.1^{+0.3}_{-0.3}$,
$\rH=0.17^{+0.17}_{-0.1}h^3\ru$,
$\dL=-2.2^{+0.4}_{-\infty}$,
$\Lmax=3.5^{+3.5}_{-1.6}\times 10^{52} h^{-2} \ergs$,
with $\chisq=9.1$ for 7 degrees of freedom. The value of $\Lmin$ is not
constrained because very faint bursts can only be seen in a small
volume and so are not well sampled for $\beta>-2.5$. The redshift distribution
for GRBs is shown in the top panel of Figure 1. As one can see,
the predicted distribution peaks at $z\sim 1$, and
about 15 percent of the bursts have redshifts larger than 2.5. 
For comparison, we show in the same panel the redshift distribution 
predicted by a standard-candle model and where the burst rate is proportional
to the star formation rate (this model is not to be confused with the model
presented at the beginning of the section
where the comoving rate of GRBs is constant).
With $\chisq=17$ for 9 degrees of freedom,
this model is not favored by the data. Note that the fraction of 
high-redshift hosts in the luminosity-function model is
actually lower than that in the standard-candle model,  
because intrinsically faint bursts are numerous and can be observed
only when they are nearby.

In a standard-candle model, the peak-count rate has a one-to-one
correspondence with redshift. This is in contradiction
with the observations that GRB 971214 and 980703 have similar peak-count 
rates but very different redshifts (see below). The one-to-one 
correspondence no longer holds when we incorporate a
luminosity function for the bursts. This is illustrated 
in the bottom panel of Fig. 1 where we plot the
predicted redshift distributions for bursts in two count-rate ranges, 
$1<P<3$ (solid) and $P>8$ (dashed). For the standard-candle model,
all bursts with $P>8$ have $z<1.3$ while those with $1<P<3$ 
have $z>2.2$. The three vertical ticks indicate the redshifts
of GRB 970508 ($z=0.835$, Bloom et al. 1998), GRB 971214 
($z=3.42$, Kulkarni et al. 1998), and GRB 980703 ($z=0.966$,
Djorgovski et al. 1998a). These bursts have $P=0.96, 1.95, 2.4$ respectively. 
The shaded region indicates the probable redshift range, $1.5\la z
\la 2.5$
for GRB 970228 which has $P=9$ (Van Paradijs et al. 1997). As one can see,
the observed redshifts can well be accommodated in the
luminosity-function model but probably not in the 
standard-candle model. We caution, however, that
the lower cutoff in redshift for the $1<P<3$ bursts
is sensitive to the width of the luminosity function, $\dL$,
which is not well constrained by the present data.

\begin{figure}
\vspace{-1cm}
\resizebox{\hsize}{!}{\includegraphics{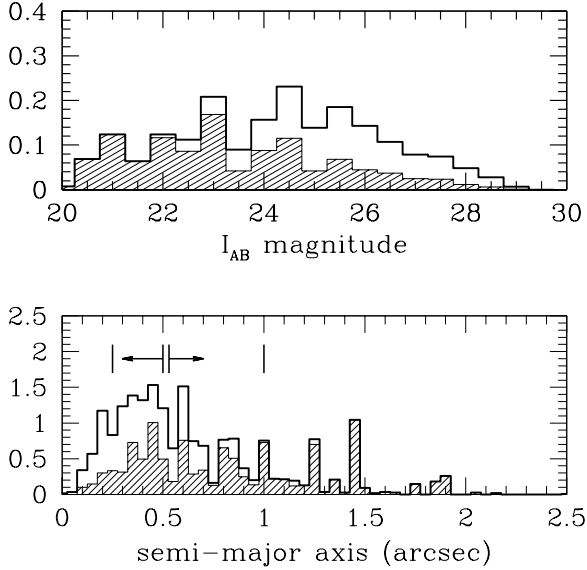}}
\vspace{-0.8cm}
\caption{
The predicted magnitude (top panel)
and semi-major axis size (bottom panel) distributions for host galaxies of
GRBs. The shaded histograms are for GRBs with $z<1.5$.
The size or size limits for four GRB hosts are indicated (see text).
}
\end{figure}

One striking feature of the observed GRB hosts is that they
have very faint magnitudes and small sizes. 
This lack of bright GRB hosts was called the ``no host'' problem 
before the discovery of the optical counterparts (Schaefer 1998). 
Here we examine whether this feature can be explained in our model.
To make theoretical predictions for the size and magnitudes of the 
host galaxies, it is necessary to know how star formation is
partitioned in galaxies with different sizes and luminosities. 
Unfortunately, this information is not yet complete,
particularly in the redshift range from 1.5 to 2.5 (where there are
no optical lines for identifying redshifts). The situation 
will be improved in the future by the use of high-resolution 
infrared spectrographs. At the moment, however, one has to
make assumptions based on the number counts of faint galaxies 
(see Hogg \& Fruchter 1998) or use photometric redshifts. 

In this paper, we adopt the second approach. We
use the photometric redshift sample of Lanzetta et al. (1996) 
in the Hubble Deep Field (Williams et al. 1996) to sample
the properties of the GRB host galaxies.
The Lanzetta et al. sample provides the I-band magnitude (in the AB system),
semi-major axis sizes and photometric redshifts for 1683 galaxies. We
use the tabulated flux at 3000\AA~ rest (calculated from the template
spectra) as an indicator for the star formation rate (Lilly et al. 1996).
The photometric redshifts are reasonably accurate (Hogg et al. 1998),
and are 
sufficient for our purpose of assigning an approximate redshift
to a GRB host galaxy in a statistical sense. A Monte-Carlo approach
is adopted to assign size and luminosity to a host galaxy.
We first generate the redshift of a burst according to the redshift 
distribution shown in the top panel of Fig. 1.
We then select a galaxy randomly from the galaxies in the Hubble Deep Field
that are within $\delta z=0.04$ of the redshift generated,
with the probability of choosing a particular galaxy 
being proportional to its star formation rate.
Fig. 2 shows the resulting magnitude and size distributions
for the GRB hosts. The median $I_{AB}$ magnitude is about 24.5 magnitude,
with 90 percent of the galaxies lying in the range from
21.5 to 28 magnitude. This distribution is quite similar to that
obtained by Hogg \& Fruchter (1998) using different procedures.
Seven of the nine observed GRB host galaxies have Vega-calibrated
$R_{AB}$ magnitudes between 24.5-25.7 (Hogg \& Fruchter 1998), 
roughly corresponding to our $I_{AB}$ of 24.0-25.2. These galaxies
are therefore at the peak of the magnitude distritution. For the other two 
galaxies, one has $I_{AB} \sim 22$, and one has $R_{AB}>22$. Clearly, a
larger sample of GRB host galaxies is needed to see if the
predicted magnitude distribution indeed matches the observed one.
The expected size are quite small, with the median semi-major axis being
about $0.5\arcsec$ and 90 percent of the hosts having sizes smaller than
1.3\arcsec. The shaded histogram shows the magnitudes and sizes of 
host galaxies with $z<1.5$. Not surprisingly, these galaxies are on
average brighter and have larger sizes than the hosts at
$z>1.5$. Notice
that a fair fraction of galaxies at $z<1.5$ are also faint and small.
These are the faint blue galaxies which dominate the number 
counts and may have a substantial contribution to the star 
formation rate at intermediate redshifts (see Ellis 1997 
for a review). For comparison, we show the sizes or size limits for four 
observed galaxies, GRB 970228 ($\sim 1\arcsec$, Sahu et al. 1997),
GRB 970508 ($\approx 0.25\arcsec$, Fruchter et al. 1998), GRB 971214
($\ga 0.5\arcsec$, Kulkarni et al. 1998), GRB 980703 ($\la 0.5$\arcsec,
Djorgovski et al. 1998a). The predicted size distribution is consistent
with the observations.

\section{Summary and Discussion}

We have studied the properties of the GRB hosts in
a scenario where the burst rate is
proportional to the star formation rate and the effect 
of the burst luminosity function is taken into account. 
The GRB hosts have their redshift distribution peaked around
$z\sim 1$, and about 15 percent have $z>2.5$. Since the star formation
rate at $z\sim 3$ is dominated by Lyman-break galaxies 
(Steidel et al. 1998), this high-redshift fraction of hosts 
should have properties similar to that of the Lyman-break galaxies. 
It is therefore interesting to note that the host
of GRB 971214 indeed resembles a Lyman-break galaxy found
at comparable redshift (Kulkarni et al. 1998). These high-redshift
host galaxies
likely have circular velocity larger than $250\kms$ (Mo, Mao \& White
1998b), while those lower redshift hosts (in particular the faint blue
galaxies) may have smaller 
circular velocity ($\sim 50-100\kms$). The difference in the
circular velocity may be relevant for distinguishing 
the ``hypernova'' model from the scenario where GRBs are
produced by mergers of binary neutron stars (Bloom et al. 1998).
The sizes of GRB hosts are small, 
90 percent of them having sizes smaller than 1.3\arcsec. The observed sizes
match this prediction. The host galaxies are faint and have
$I_{AB}$ between 21.5 and 28.
Most host galaxies are within one magnitude of the predicted
most-likely value, and there seems to be a lack of bright hosts
compared with the prediction. This may, however,
be due to a selection effect: bright galaxies presumably are more 
metal-rich and so the GRB afterglows may suffer more dust extinction.
More observations of GRB host galaxies are needed to give a stringent 
constraint on the model.

Although our best-fit model has a luminosity width of about two decades, most
bursts occur in a narrower range. This is because the
faint bursts can be observed only
locally while the bright ones are not
numerous. For the best fit, the median peak
luminosity is $\approx \Lstar = \Lmax/30 \approx 10^{51}h^{-2}\ergs$,
while 90\% of GRBs are within $\Lstar/3<L<8\Lstar$.
The ``effective'' duration for the long GRBs is 
$\Delta t \approx 10{\rm s}$
(Mao, Narayan \& Piran 1994), therefore the typical total energy of GRBs is 
$\approx \Lstar \Delta t = 10^{52}$ $h^{-2}$ $\ergs$. For
a flat model with $\Omega_0=0.3, \Lambda_0=0.7$, both the peak luminosity
and total energy are larger by a factor of 2.5. Note, however, that the
{\it maximum} peak luminosity and total energy can be a factor of $\sim
10$ larger.

Further tests of the model come from gravitational lensing
and the cosmological time dilation of GRBs. To estimate the lensing
probability, we model galaxies as
singular isothermal spheres with constant comoving 
number density. The lensing probability is about one 
in two thousand, and therefore the number of lensing events 
in the BATSE experiment should not be significant.
The predicted relative time dilation for bursts with $P>8$ and $1<P<3$ 
is about a factor of 1.3, consistent with the lower end of the 
values reported by Bonnell et al. (1997). Note that
such analyses make the implicit assumption that bursts 
at different redshifts are statistically the same, which
may not be true. For example, 
in our model which takes into account the burst luminosity function, 
intrinsically faint bursts mostly occur at low
redshift (cf. Fig. 1). 
So some caution should be exercised in interpreting results which are
based on the standard candle model (e.g., Deng \& Schaefer 1998).
If the GRB duration is luminosity-dependent,
then the interpretation of the time dilations will be more complicated.
In addition,
galaxies themselves evolve, for example, galaxies at high-redshift are 
smaller and denser (e.g. Mo, Mao \& White 1998a). Such evolution
might affect the predictions of GRB afterglows since
they all depend on the density of the ambient medium (e.g., Waxman 1997).

\begin{acknowledgements}
We are grateful to Bohdan Paczy\'nski for encouragement, 
to him, Peter Schneider and Simon White for helpful
comments on the paper, to A. Fenandez-Solo for information
about HST data.


\end{acknowledgements}

\vspace{-0.5cm}


\end{document}